\documentclass[journal]{IEEEtran}

\ifCLASSINFOpdf
\else
   \usepackage[dvips]{graphicx}
\fi
\usepackage{url}
\usepackage{cite}
\usepackage{amssymb,amsfonts}
\usepackage{algorithmic}
\usepackage{textcomp}
\usepackage{xcolor}
\usepackage{stfloats}
\usepackage{float}
\usepackage{subfig}
\usepackage{booktabs}
\usepackage{mathtools}
\usepackage[colorlinks=true]{hyperref}
\usepackage{siunitx}
\usepackage{upgreek}
\usepackage{caption}
\usepackage{graphicx}

\begin{document}

\title{Deep Learning-Based Approach for Identification and Compensation of Nonlinear Distortions in Parametric Array Loudspeakers}

\author{Mengtong Li, Tao Zhuang, Kai Chen, \IEEEmembership{Member, IEEE}, Jia-Xin Zhong, \IEEEmembership{Member, IEEE}, and Jing Lu, \IEEEmembership{Member, IEEE}
\thanks{Manuscript received xxxxxxx xx, 2024; revised xxxxxxx xx, 2024; accepted xxxxxxx xx, 2024. Date of publication xxx xx, 2024; date of current version xxx xx, 2024. This work was supported by the National Natural Science Foundation of China under Grant 12274221. The associate editor coordinating the review of this manuscript and approving it for publication was xxx. (Corresponding author: Jia-Xin Zhong.)}
\thanks{Mengtong Li, Tao Zhuang and Jing Lu are with the Key Laboratory of Modern Acoustics, Nanjing University, Nanjing 210008, China, and also with NJU-Horizon Intelligent Audio Lab, Horizon Robotics, Beijing 100094, China (e-mail: mengtong.li@smail.nju.edu.cn; taozhuang@smail.nju.edu.cn; lujing@nju.edu.cn).}
\thanks{Kai Chen is with the Key Laboratory of Modern Acoustics, Nanjing University, Nanjing 210008, China (e-mail: chenkai@nju.edu.cn).}
\thanks{Jia-Xin Zhong is with the Graduate Program in Acoustics, The Penn sylvania State University, University Park, PA 16802 USA (e-mail: jiaxin.zhong@psu.edu).}}

\markboth{Journal of \LaTeX\ Class Files, Vol. 14, No. 8, August 2015}
{Shell \MakeLowercase{\textit{et al.}}: Bare Demo of IEEEtran.cls for IEEE Journals}
\maketitle

\begin{abstract}
Compared to traditional electrodynamic loudspeakers, the parametric array loudspeaker (PAL) offers exceptional directivity for audio applications but suffers from significant nonlinear distortions due to its inherent intricate demodulation process.
The Volterra filter-based approaches have been widely used to reduce these distortions, but the effectiveness is limited by its inverse filter's capability. 
Specifically, its $p$th-order inverse filter can only compensate for nonlinearities up to the $p$th order, while the higher-order nonlinearities it introduces continue to generate lower-order harmonics. 
In contrast, this paper introduces the modern deep learning methods for the first time to address nonlinear identification and compensation for PAL systems. 
Specifically, a feedforward variant of the WaveNet neural network, recognized for its success in audio nonlinear system modeling, is utilized to identify and compensate for distortions in a double sideband amplitude modulation-based PAL system.
Experimental measurements from 250\,Hz to 8\,kHz demonstrate that our proposed approach significantly reduces both total harmonic distortion and intermodulation distortion of audio sound generated by PALs, achieving average reductions to 4.55\% and 2.47\%, respectively. 
This performance is notably superior to results obtained using the current state-of-the-art Volterra filter-based methods.
Our work opens new possibilities for improving the sound reproduction performance of PALs.

\end{abstract}


\IEEEpeerreviewmaketitle

\section{Introduction}

\IEEEPARstart{D}{ue} to their exceptional capacity for generating highly directional audio sound beams through the nonlinear interactions between intensive ultrasound, parametric array loudspeakers (PALs) have gained considerable attention in various audio applications \cite{bookAWPAL,Gan2012review}, such as sound reproduction \cite{shi2014overview}, active noise control \cite{Zhong2022QZ} and omnidirectional sound sources \cite{Arnela2022}. 
When a PAL radiates an ultrasound carrier modulated by an audio signal, the audio signal will be self-demodulated during the propagation in the air due to the second-order nonlinearity. 
However, because of the inherent nonlinearity, the audio sound generated by PALs suffers from significant nonlinear distortion, which limits their broader application in audio engineering.

\begin{figure}[t]
\vspace{3mm}
\centerline{\includegraphics[width=0.5\textwidth]{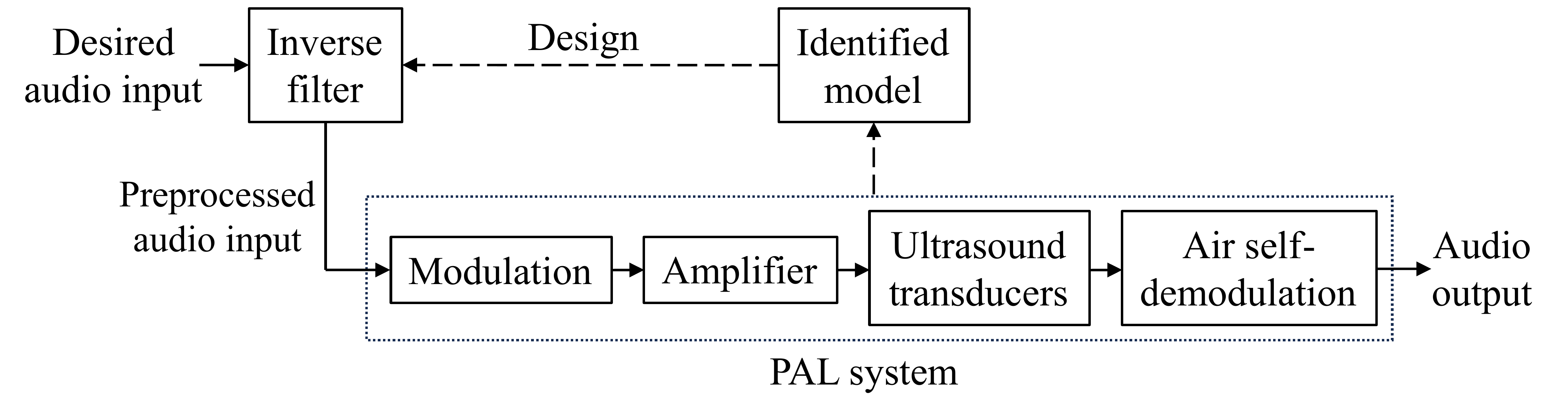}}
\caption{PAL compensation using an inverse filter.}
\vspace{-3mm}
\label{fig1}
\end{figure}

To reduce the nonlinear distortion, early studies focused on developing various modulation methods to compensate for the nonlinearity \cite{Kamakura1984SRAM,Kamakura1985SSBAM,Tan2010MAM_IMD}. 
However, these methods are based on a physical law-rooted mathematical models called Berktay's far-field solution \cite{Berktay1965}, which relies on various assumptions, including weak nonlinearity, collimated beams, ideal flat-band frequency response of transducers, far-field observation points, and so on. 
These assumptions are challenging to meet in real-world scenarios, resulting in a significant decline in the distortion reduction performance. 

A more effective approach to mitigate the nonlinear distortion, as schematically presented in Fig.~\ref{fig1}, involves identifying the whole PAL process including modulation, transducer's frequency response and airborne self-demodulation as a nonlinear audio system, and subsequently designing an inverse filter to preprocess the input signals, thereby compensating for nonlinearity. 
There are various models for nonlinear system identification, among which the Volterra filter (VF) has been widely used for PAL system identification and compensation \cite{Lee2002VF_AES,Ji2012inverse,Mu2014ODVF,Shi2015nlms,Shi2016JASA,Hatano2017TASLP}. 
VF adopts a straightforward filter structure to model the system's nonlinearity, expressing the input-output relationship as a polynomial series.
The least mean squares (LMS) algorithms or the frequency response method can be adopted to obtain the coefficients of the VF, which are then be used to design an inverse filter structure to compensate the system's nonlinearity.
However, the number of coefficients grows exponentially with the order of VF, and the approximation rate of this polynomial structure suffers from the ``curse of dimensionality'' \cite{Barron1993Universal}, making higher-order VFs challenging to determine. 
More importantly, while compensation based on a $p$th-order inverse filter can reduce distortion, its effectiveness is limited. 
This limitation arises because the $p$th-order inverse structure of VF cancels out lower-order nonlinearities by introducing higher order nonlinearities, yet higher order nonlinearities still generate lower harmonics. 
For instance, all even-order nonlinearities will produce the second harmonic, making it significantly challenging to reduce harmonic distortions to a low level. 

Recently, the application of deep learning in system identification has attracted intensive interest due to its powerful nonlinear fitting ability and high flexibility that could adapt to highly complex dynamical systems. 
In the realm of nonlinear audio distortion effects modeling, deep learning approaches have shown remarkable performance \cite{Damskagg2019AaltoICASSP,Nercessian2021LightweightICASSP,Sudholt2023PruningTASLP}. 
With deep neural networks, it is possible to identify PAL systems and fit a complete inverse, rather than being limited to a $p$th-order inverse.
Despite this potential, research on applying deep learning to PAL identification and compensation remains scarce, except for \cite{Zhou2022NNinv}, which employed a simple 3-layer fully-connected network. 
However, such a preliminary model is hampered by its limited capacity to represent the complex nonlinearity of PALs, resulting in unsatisfactory performance. 
Moreover, the study does not use any metric to assess the reduction of nonlinear distortion, and based on the spectrograms they presented, the method appears ineffective at mitigating nonlinear distortion. 

This paper focuses on a modern model based on WaveNet convolutional neural network \cite{oord2016wavenet}, which has been successfully used for modeling the nonlinearity in vacuum-tube amplifier \cite{Damskagg2019AaltoICASSP}.
The model is made up of a stack of dilated causal convolution layers, which is effective for nonlinear modeling in time domain. 
In this study, the model is applied to both PAL identification and compensation. 
Performance is evaluated and compared to the state-of-the-art VF-based approaches using total harmonic distortion (THD) and intermodulation distortion (IMD), two typical metrics commonly used for assessing nonlinear distortion. 

\section{Deep learning-based identification and compensation}
\begin{figure}[t]
\centerline{\includegraphics[width=0.5\textwidth]{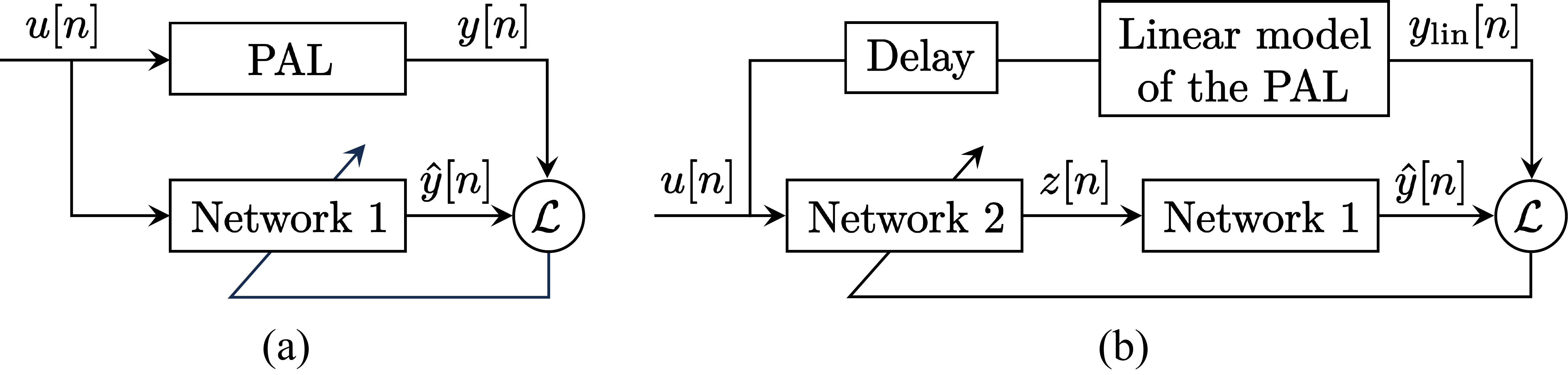}}
\caption{Deep learning-based nonlinear system identification and compensation for a PAL.
 (a) Identified model training, (b) inverse filter training for compensation. $\mathcal{L}$ represents the loss function. }
\vspace{-3mm}
\label{fig2}
\end{figure}
Figure~\ref{fig2} illustrates the two-step procedure for identification and compensation for a PAL system using deep learning. 
In the first step, as illustrated in Fig.~\ref{fig2}(a),  PAL system identification, Network 1 is trained to approximate the input–output behavior of the PAL system from measured data, serving as the identified model in Fig.~\ref{fig1}. 
The identified model is required to predict the current output sample $y[n]$ based on a given sequence of past input samples along with the current input sample.
Thus, Network 1 represents the map 
\begin{equation}
    \hat{y}[n,\boldsymbol{\theta}] = f(u[n],...,u[n-N+1],\boldsymbol{\theta}),
    \label{map1}
\end{equation}
where $u[n]$ and $\hat{y}[n]$ are respectively the input and the estimated output at time $n$, $N$ represents the receptive field of the model, and $\boldsymbol{\theta}$ represents the parameters of Network 1. 
The model is trained by minimizing the discrepancy between the estimated output $\hat{y}[n]$ and the measured output $y[n]$.

Upon obtaining the identified model (Network 1), as shown in Fig.~\ref{fig2}(b), the next step introduces Network 2 to implement the inverse filter, with the desired signal $u[n]$ as its input and the preprocessed signal $z[n]$ as output. 
To maintain the preprocessed signal $z[n]$ within the input dynamic range of the actual PAL system, an amplitude constraint is applied to the output of Network 2.
Using the similar representation as Network 1, the inverse filter is represented as the map 
\begin{equation}
    z[n,\boldsymbol{\varphi}] = g(u[n],...,u[n-N+1],\boldsymbol{\varphi}),
    \label{map2}
\end{equation}
where $\boldsymbol{\varphi}$ is its parameters. 
By passing the preprocessed signal $z[n]$ through the PAL system simulated by Network 1, the estimated system output $\hat{y}[n]$ is obtained. 
The output signal of the PAL system can be expressed as $y[n] = y_\mathrm{lin}[n] + y_\mathrm{nlin}[n]$, where $y_\mathrm{lin}[n]$ and and $y_\mathrm{nlin}[n]$ represent the linear and nonlinear components of $y[n]$, respectively. 
The compensation of nonlinear distortions is to eliminate the nonlinear component $y_\mathrm{nlin}[n]$ while retaining the linear component $y_\mathrm{lin}[n]$. 
Thus, Network 2 is trained to minimize the discrepancy between $\hat{y}[n]$ and $y_\mathrm{lin}[n]$ with a time delay, where $y_\mathrm{lin}[n]$ is derived from a linear PAL model identified using a finite impulse response (FIR) filter via the LMS algorithm.
According to the mechanism of linear inverse systems, a time delay is necessary for the inverse system of a non-minimum phase system to remain stable. 
Since systems involving acoustic signal transmission, such as PALs, are generally considered non-minimum phase systems \cite{Neely1979linInv}, a time delay is introduced here to ensure stability and causality of the inverse system. 

\begin{figure}[t]
\centerline{\includegraphics[width=0.38\textwidth]{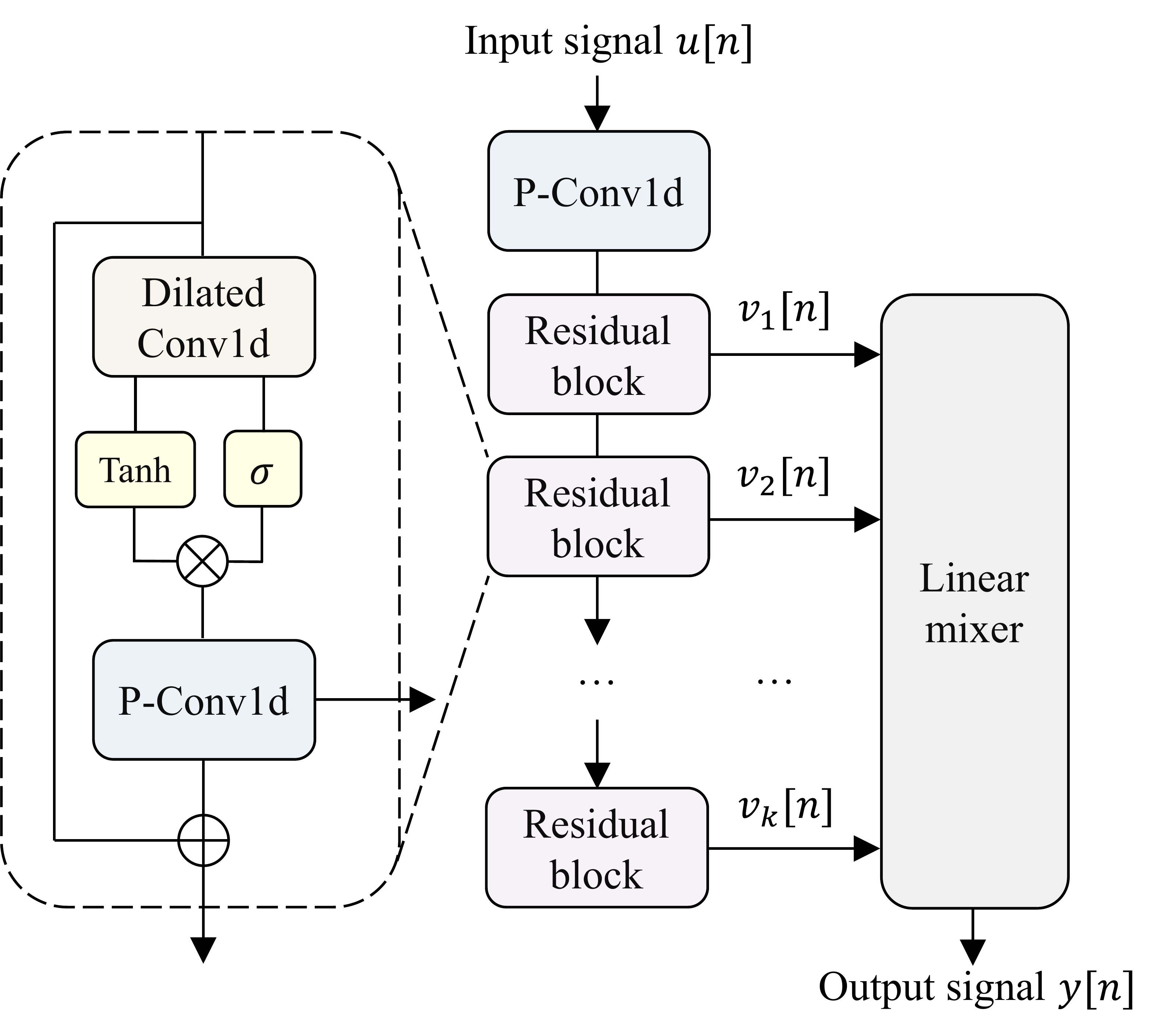}}
\caption{The structure of the feedforward WaveNet.}
\vspace{-3mm}
\label{fig3}
\end{figure}

In this work, a feedforward variant of the WaveNet architecture \cite{Damskagg2019AaltoICASSP} is employed to fit the maps (\ref{map1}) and (\ref{map2}) in both Network 1 and 2. 
As shown in Fig.~\ref{fig3}, the input and output of the model are signal sequences. 
Through the first pointwise (kernel size = 1) 1D convolution layer (P-Conv1d), the channel number of the input signal $u[n]$ is adjusted from 1 to $C$. 
The adjusted signal then passes through a series of residual blocks, each comprising a 1D dilated convolution layer (Dilated Conv1d), a gated Tanh activation and a pointwise 1D convolution layer.
Additionally, a residual connection is employed where the input of the block is added to the output of the activation, yielding the block's final output. 
The outputs of the residual blocks are stacked and then fed into a ``linear mixer'', a pointwise 1D convolution layer, to produce the final output sequence.
Stacked dilated convolutions allows the network to achieve a large receptive field with just a few layers, while maintaining input resolution and computational efficiency. 
The receptive field $N$ of the model is given by $N = (M-1)\sum_{k=1}^K d_k + 1$, where $M$ is the kernel size of the 1D dilated convolution layers, $K$ is the number of residual blocks, and $d_k$ is the dilation factor of the $k$-th 1D dilated convolution layer.

The loss function used in both the identification and inverse filter training processes is applied on both the waveform and spectrogram domains:
\begin{equation}
    \mathcal{L} = \mathrm{MSE}(y_1,y_2) + \mathrm{MSE}(|Y_1|,|Y_2|),
    \label{loss}
\end{equation}
where $y_1$ and $y_2$ are the target waveform and the model output waveform, and $Y_1$ and $Y_2$ are their spectrogram. 
In Network 1 (2), $y_1$ and $y_2$ represent $y$ ($y_\mathrm{lin}$) and $\hat{y}$ ($\hat{y}$), respectively. 

\section{Experiments}
\subsection{Experimental setup}
\begin{figure}[t]
\centerline{\includegraphics[width=0.48\textwidth]{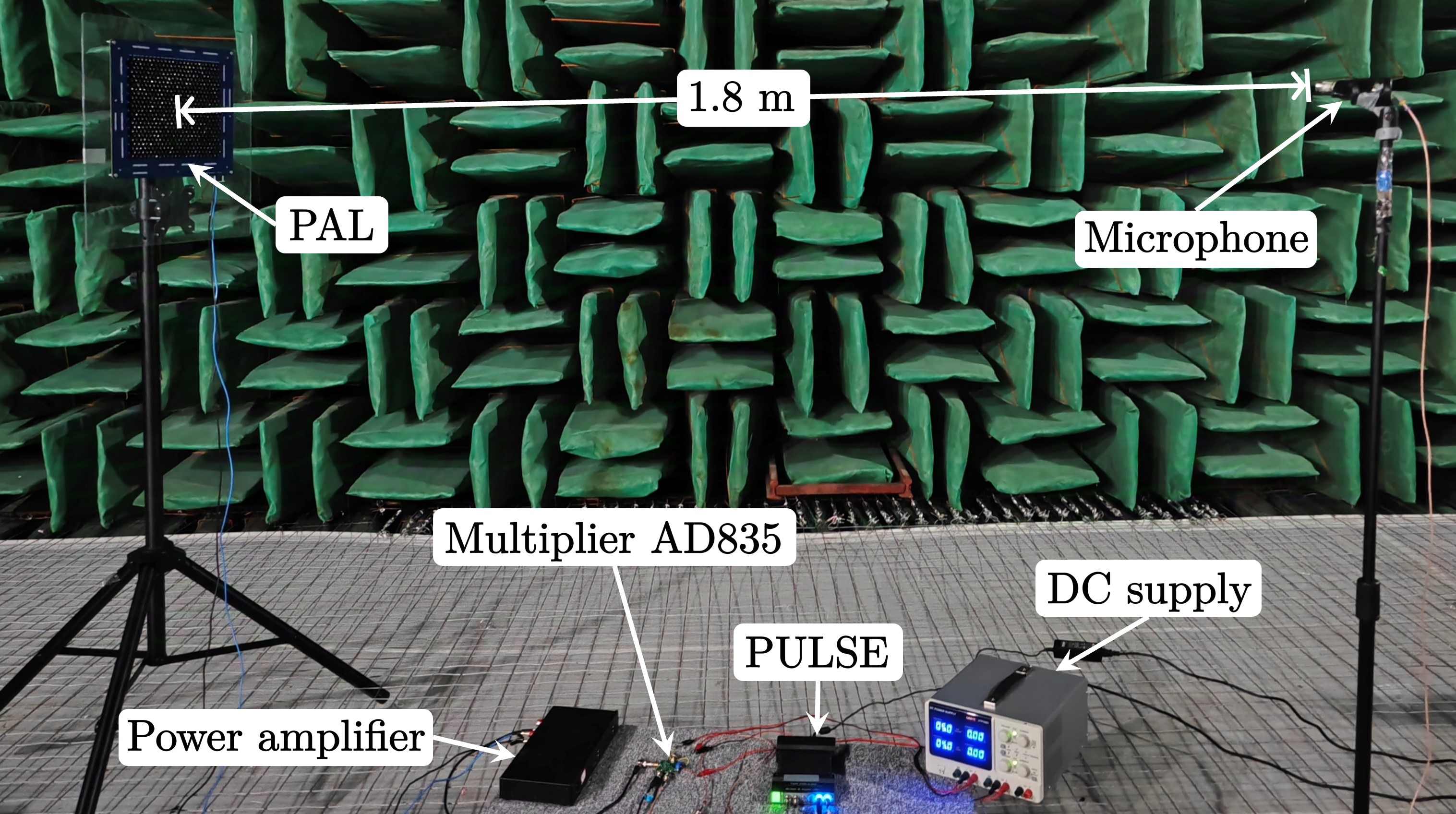}}
\caption{A photo of the experimental setup.}
\label{expfig}
\end{figure}
\begin{figure}[t]
\centerline{\includegraphics[width=0.48\textwidth]{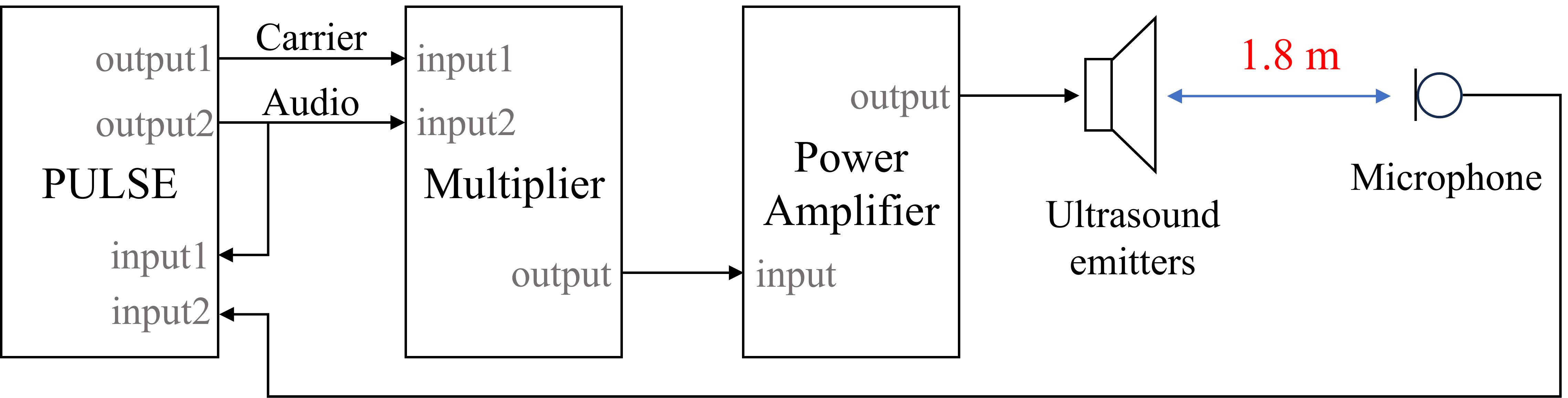}}
\caption{The signal flow of the measurement.}
\vspace{-3mm}
\label{expsignalfig}
\end{figure}
As shown in Fig.~\ref{expfig}, the experiments are carried out at a full anechoic chamber in Nanjing University with dimensions of $\SI{11.4}{m}\times \SI{7.8}{m} \times \SI{6.7}{m}$. 
The relative humidity and temperature are $\SI{48}{\%}$ and $\SI{26}{^\circ C}$, respectively. 
The laboratory-fabricated PAL prototype used in experiment consists of 256 circular ultrasonic emitters (Murata MA40S4S, Kyoto, Japan) which has a resonance frequency of $\SI{40}{kHz}$ and a radius of $\SI{5}{mm}$. 
The ultrasonic emitters are arranged in a compact square array with side lengths of around $\SI{16}{cm}$. 
Figure \ref{expsignalfig} shows the signal flow of the measurement. 
The audio input and the ultrasound carrier of the PAL are generated by a B\&K PULSE system (Type 3160).
The modulation method used in the system is DSBAM (double sideband amplitude modulation) \cite{Yoneyama1983DSBAM}, which is implemented by an analog multiplier Analog Devices AD835 with the modulation index of 0.9. 
The modulated signal is then fed to the ultrasonic emitter array through a power amplifier, where the peak level of voltage on the emitter is about $\SI{9}{V}$. 
The audible sound generated by the PAL is captured by a microphone at a distance of $\SI{1.8}{m}$ from the PAL and recorded by the 3160 PULSE system. 
The sampling rate of the PULSE system is $\SI{131.072}{kHz}$, and the signals recorded are resampled to $\SI{44.1}{kHz}$. 
To avoid spurious sound induced by the intensive ultrasound \cite{Ji2019spurious}, the microphone is covered by a piece of small and thin plastic film, which could reduce the ultrasound by $\SI{30}{dB}$. 

For dataset recording for PAL identification, the input audio signal of the PAL is simultaneously recorded along with the audible sound generated by the PAL. 
The audio signals used as input to the PAL are obtained from an audio database released for general sound event detection \cite{trowitzsch2019nigens}. 
This database consists of multiple classes of sound, including music, speech and a variety of ambient sounds, covering a wide range of scenarios in sound reproduction.
The experiment recorded 2 hours of input and output data to serve as the dataset for PAL system identification, which are split into 65536-sample segments for batch training. 
For inverse filter training, only system input signals are required. 
In this work, the input portion of the dataset for identification is used to train the inverse filter. 

\subsection{Evaluation metrics}

THD and IMD stand as two typical evaluation metrics for loudspeaker nonlinear distortion.
The THD is calculated by comparing the power sum of the harmonic components to the total input power, which includes both the fundamental and distortion components \cite{IEC_THD}. 
The definition of IMD in this paper follows the definition in \cite{Tan2010MAM_IMD}, which is calculated by comparing the power sum of the intermodulation components to the total power of two fundamental components and the intermodulation components. 
A step sine is used as the THD testing audio input, of which the frequency varies across the center frequencies of one-third octave bands from $\SI{250}{Hz}$ to $\SI{8}{kHz}$. 
In the IMD measurement, another pure tone at $\SI{1.7}{kHz}$ is added in, and the amplitude of the pure tone is 4 times as that of the step sine as specificed in \cite{Shi2016JASA}. 
The THD and IMD are calculated up to the 4th order in this study.

\subsection{Nonlinear identification}
The WaveNet model used for identification has 9 residual blocks, where the dilated convolution layers have a dilation pattern of $d_k=\{1, 2, 4, ..., 256\}$, and share a common channel number $C=16$ and a common kernel size $M=16$ with the pointwise convolution layers. 
The model is trained by Adam Optimizer with a batch size of 8. 
The initial learning rate is $0.001$ and it will be reduced by the factor of 0.1 if the validation loss does not decrease for 10 consecutive epochs. 
Gradient clipping with a maximum norm of 5 is used.

\begin{figure}[t]
\centerline{\includegraphics[width=0.49\textwidth]{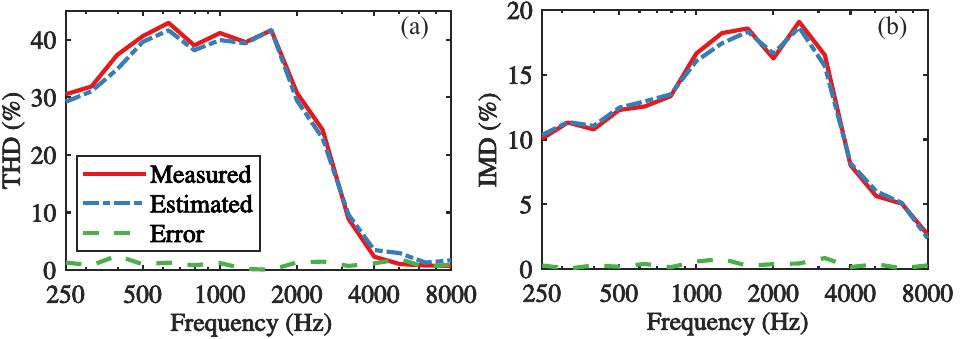}}
\caption{The results of PAL system identification using WaveNet. (a) Measured and model-estimated THD and their error. (b) Measured and model-estimated IMD and their error. The same legend in (a) applies to (b).}
\vspace{-2mm}
\label{fig5}
\end{figure}

To assess the accuracy of the identified model, the THD and IMD estimated by the model are compared with that measured by experiment, as depicted in Fig.~\ref{fig5}. 
The results show that the estimated THD and IMD closely match the measured values, with only slight discrepancies. 
Specifically, the average error in THD and IMD estimation are only 1.08 \% and 0.34 \%, respectively. 
Thus the model is considered to be ready for the PAL system simulation in the next step.

\subsection{Nonlinear compensation}
The model used for compensation has 24 residual blocks, where the dilated convolution layers have a dilation pattern of $d_k=\{1, 2, 4, ..., 2048, 1, 2, 4, ..., 2048\}$, and share a common channel number $C=24$ and a common kernel size $M=4$ with the pointwise convolution layers. 
Additionally, a Tanh activation is added after the linear mixer to limit the output amplitude of the model. 
The time delay introduced in inverse filter is set to 100 samples (about $\SI{2.3}{ms}$). 
The model is trained by Adam Optimizer with a batch size of 8. 
The initial learning rate is 0.001 and it will be reduced by the factor of 0.2 if the validation loss does not decrease for 5 consecutive epochs. 
Gradient clipping with a maximum norm of 5 is also used.

For comparison, the VF-based method is also implemented in the experiment. 
The input signal used for identifying the VF is a white Gaussian noise signal with a time duration of $\SI{45}{s}$. 
The normalized least mean squares (NLMS) algorithm was adopted to estimate the 2nd-order truncated VF, with the auxiliary step size of 0.01. 
The memory length of the 1st- and 2nd-order VF kernels are 160 and 80, respectively. 
The 2nd-order and 3rd-order inverse filter based on the identified VF kernels follows the structure introduced in \cite{bookVolterraWiener}, where the time delay is also set to 100 samples as described above. 
Both deep learning-based and VF-based inverse filter are used for offline processing. 

\begin{figure}[t]
\centering
\captionsetup[subfloat]{labelformat=empty}
\subfloat[]{\includegraphics[width=0.43\textwidth]{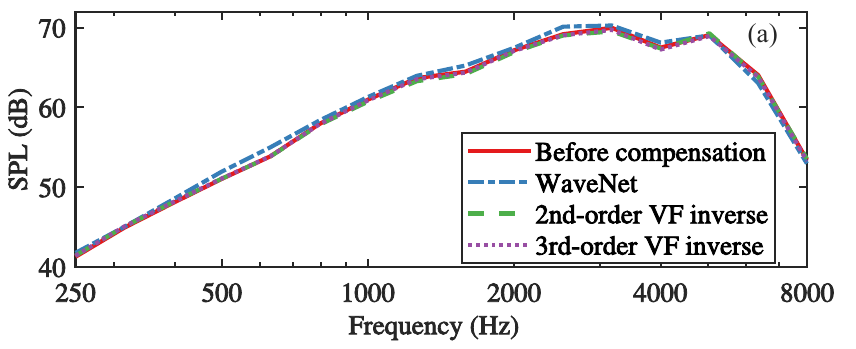}
\label{fig_61}}
\vspace{-6mm}
\\
\subfloat[]{\includegraphics[width=0.43\textwidth]{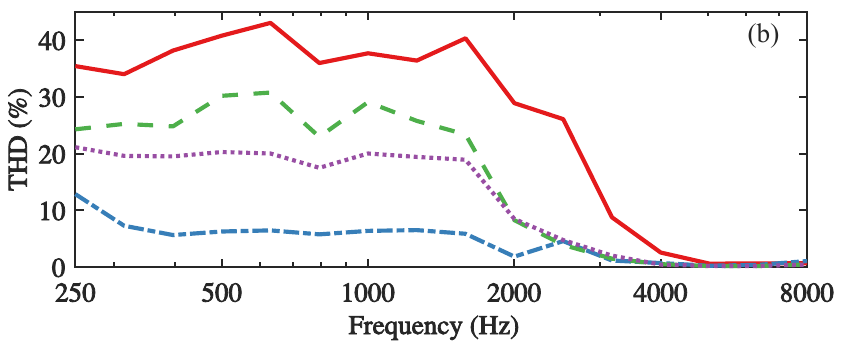}
\label{fig_62}}
\vspace{-6mm}
\\
\subfloat[]{\includegraphics[width=0.43\textwidth]{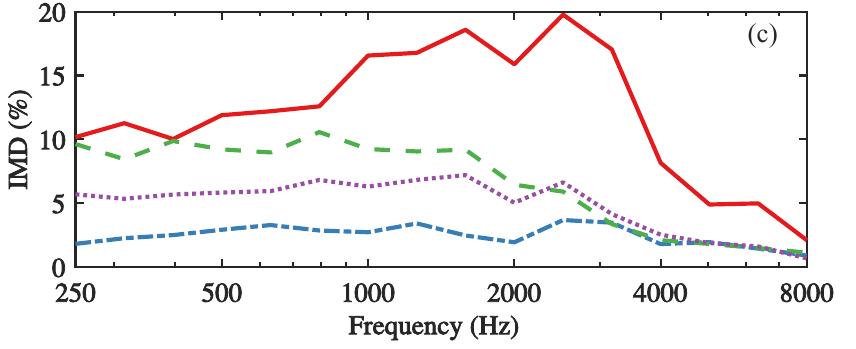}
\label{fig_63}}
\vspace{-3mm}
\caption{Results measured before and after compensation by inverse filters. (a) the linear responses, (b) the THD, (c) the IMD. The same legend in (a) applies to all plots.}
\label{fig6}
\end{figure}

For the PAL system before and after compensation, the measured linear responses are presented in Fig.~\ref{fig_61}. 
It is noted that both the VF-based and the deep learning-based inverse filters have minimal effects on the linear response of the PAL system. 
Figure~\ref{fig_62} and Figure~\ref{fig_63} show the measured THD and IMD of the PAL before and after compensation. 
For the convenience of comparison, the average of the distortion before and after the compensation using deep learning-based method and the VF-based method are listed in Table~\ref{table1}. 
The table also lists the average THD and IMD from $\SI{250}{Hz}$ to $\SI{4}{kHz}$, the frequency range where the major energy of speech and music signals is concentrated. 
The results show that deep learning-based compensation effectively reduces the average THD to below 5\%, a significant improvement over the VF-based methods. 
For intermodulation distortion, the deep learning-based compensation reduces the IMD to below 3\%. 
Notably, the deep learning-based method outperforms the VF-based method across all frequencies. 


\begin{table}[t]
    \centering
    \caption{The average of the distortion before and after the compensation. 
    ``VF2'' and ``VF3'' represent the 2nd-order and 3rd-order inverse of VF, respectively.}
    \begin{tabular}{ccccc}
    \toprule
         & Before & WaveNet & VF2 & VF3 \\ 
        \midrule
        THD ($\SI{250}{Hz}$-$\SI{8}{kHz}$) & 25.62  &  4.55  &   15.70   &   12.04 \\ 
        IMD ($\SI{250}{Hz}$-$\SI{8}{kHz}$) & 12.05  &  2.47  &   6.65      &   4.87 \\ 
        THD ($\SI{250}{Hz}$-$\SI{4}{kHz}$) & 31.39  &  5.47  &   19.27   &   14.76 \\ 
        IMD ($\SI{250}{Hz}$-$\SI{4}{kHz}$) & 13.91  &  2.71  &   7.85      &   5.69 \\ 
        \bottomrule
    \end{tabular}
\label{table1}
\end{table}

\section{Conclusion}
For reduction of the nonlinear distortion of the PAL, the modern deep learning methods are introduced to PAL nonlinear identification and compensation. 
In this paper, a feedforward variant of the WaveNet deep neural network is employed to model the nonlinearity of the PAL system, and is then used to compensate the nonlinearity as an inverse filter. 
Experimental results demonstrate that our proposed deep learning-based method outperforms the state-of-the-art VF-based method, offering significant potential for improving the sound reproduction performance of PAL systems.

\vfill\pagebreak

\bibliographystyle{IEEEtran}
\bibliography{IEEEabrv,reference}

\end{document}